\documentclass[prd,twocolumn,showpacs,superscriptaddress,nofootinbib]{revtex4-1}
\usepackage[T1]{fontenc} 
\usepackage{amsmath}
\usepackage{amssymb}
\usepackage{amsfonts}
\usepackage{graphicx}
\usepackage{enumitem}
\usepackage{bm}
\usepackage{rotating}
\usepackage{hyperref}
\usepackage{xcolor}
\usepackage{array}
\usepackage{lmodern}
\usepackage[normalem]{ulem}
\usepackage{braket}
\usepackage{tensor}
\usepackage{mathrsfs}
\usepackage{float}

\usepackage[normalem]{ulem}

\begin{document}

\title{Undemocratic Dirac seesaw}

\author{Su-Ping Chen}

\email{spchen@seu.edu.cn}

\author{Pei-Hong Gu}

\email{phgu@seu.edu.cn}

\affiliation{School of Physics, Jiulonghu Campus, Southeast University, Nanjing 211189, China}

\begin{abstract}

The standard model left-handed neutrinos and several right-handed neutrinos can obtain a tiny Dirac mass matrix through their mixings with relatively heavy Dirac fermions. In this Dirac seesaw scenario, the mixings involving the left-handed neutrinos can be allowed much larger than those involving the right-handed neutrinos. This undemocratic parameter choice is attractive to phenomenology. We show that the small mixings between the heavy Dirac fermions and the right-handed neutrinos can have a common origin with the observed baryon asymmetry in the universe. We also connect the introduction of right-handed neutrinos to the existence and stability of dark matter by a new $U(1)$ gauge symmetry for dark photon or baryon-minus-lepton number. We then specify how to embed our scenario into a left-right symmetric theory or a grand unification theory.

\end{abstract}


\maketitle

\section{Introduction}

The $SU(3)_c^{}\times SU(2)_L^{} \times U(1)_Y^{}$ standard model (SM) has been tested to a very high accuracy, but it has been suffering some big challenges such as neutrino mass, baryon asymmetry and dark matter from astronomical observations and terrestrial experiments \cite{pdg2020}.

In the SM, the charged fermions can obtain their masses through the Yukawa couplings of the left-handed fermion doublets and the right-handed fermion singlets to the Higgs doublet. However, the neutrino masses can not be induced in this way because the right-handed neutrinos are absent from the SM. Even if the right-handed neutrinos are introduced by hand, their Yukawa couplings with the SM should be very small since the neutrinos are extremely light. Currently the best explanation for the tiny neutrino masses seems to be the famous seesaw mechanism \cite{minkowski1977,yanagida1979,grs1979,ms1980,mw1980,sv1980,cl1980,lsw1981,ms1981}. In various seesaw extensions \cite{minkowski1977,yanagida1979,grs1979,ms1980,mw1980,sv1980,cl1980,lsw1981,ms1981,mv1986,flhj1989,ma2006,ghsz2009,adefhv2018,adefhv2018-2,gu2019-4,gu2019-2,gu2019-7} of the SM, the neutrino mass generation is tied to certain lepton-number-violating interactions so that the neutrinos should have a Majorana nature. However, the lepton number violation and then the Majorana neutrinos are just a theoretical assumption which has not been confirmed in any experiments. So, it should be worth exploring the possibility of Dirac neutrinos \cite{rw1983,rs1984,dlrw1999,mp2002,tt2006,ap2006,gh2006,gs2007,gu2012,fm2012,ms2015,gu2016,mp2017,wh2017,hw2018,yd2017,yd2018,saad2019,jks2019,gu2019-1,gu2019-3,gu2019-8,prsv2019,gu2021}. In analogy to the conventional seesaw models for the light Majorana neutrinos, we can construct some Dirac seesaw models for the light Dirac neutrinos \cite{rw1983,rs1984,mp2002,gh2006,gs2007,gu2012,fm2012,ms2015,gu2016,mp2017,gu2019-1,gu2019-3,gu2019-8}.

To understand the cosmic baryon asymmetry, we need a dynamical baryogenesis mechanism. This requires a CPT-invariant theory of particle interactions should satisfy the Sakharov conditions \cite{sakharov1967}: (i) baryon number nonconservation, (ii) C and CP violation, (iii) departure from equilibrium. The SM fulfils all of these conditions and then accommodates an electroweak baryogenesis mechanism \cite{krs1985}. However, the SM electroweak baryogenesis can only give a baryon asymmetry far far below the observed value. The SM must be supplemented by new ingredients to realize a successful baryogenesis mechanism.

In some seesaw models, the small neutrino masses and the observed baryon asymmetry can be simultaneously explained \cite{fy1986,lpy1986,luty1991,mz1992,fps1995,fpsw1996,crv1996,pilaftsis1997,ms1998,bcst1999,chun2017}. This is the so-called leptogenesis \cite{fy1986} mechanism. In the leptogenesis scenario, the scales of generating the neutrino masses and the baryon asymmetry are tied together and determined by the mass of the same particles. The leptogenesis could not be realized at an accessible scale unless it invokes a large fine-tuning to resonantly enhance the required CP asymmetry \cite{fps1995,fpsw1996,crv1996,pilaftsis1997}. This means that a natural leptogenesis should be achieved at a high scale, which prevents the realization of a testable neutrino mass generation.

To explain the dark matter relic density in the present universe, the SM should be extended by certain elementary particle(s). The existence of the dark matter particle(s) can be motivated by other theoretical considerations such as the supersymmetric theories, where the $R$-parity guarantees the stability of the dark matter \cite{bhs2005}. Recently \cite{gu2019-3,gu2019-4,gu2019-7,gu2019-8}, it has been shown that a $U(1)_{B-L}^{}$ \cite{langacker2009} gauge symmetry can simultaneously account for the existence and stability of the dark matter, the production of the cosmic baryon asymmetry, as well as the introduction of the lepton number violation or the right-handed neutrinos for the neutrino mass generation.

In this paper we shall demonstrate a novel Dirac seesaw scenario with rich observable phenomena. This is due to an undemocratic assignment between the couplings involving the left-handed neutrinos and those involving the right-handed neutrinos. We then shall explain such undemocratic parameter choice by an additional seesaw mechanism which also accommodates a successful Dirac leptogenesis \cite{dlrw1999,mp2002}. Within this undemocratic Dirac seesaw context, we shall also consider a $U(1)$ gauge symmetry for dark photon \cite{okun1982,holdom1986,fh1991} or baryon-minus-lepton number \cite{langacker2009} to connect the introduction of the right-handed neutrinos to the existence and stability of the dark matter. Our scenario can be embedded into a left-right symmetric theory \cite{ps1974,mp1975,mp1975-2,ms1975} or a grand unification theory (GUT).

\section{Undemocratic Dirac seesaw}

We proceed the idea of undemocratic Dirac seesaw from the toy model as below,
\begin{eqnarray}
\label{lag}
\mathcal{L}&\supset& -y \bar{l}_L^{} \phi N_R^{} - \hat{m}_N^{} \bar{N}_L^{} N_R^{} - m_R^{} \bar{N}_L^{} \nu_R^{} +\textrm{H.c.}\,. 
\end{eqnarray}
Here $l_L^{}$ and $\phi$ are the SM lepton and Higgs doublets, $\nu_R^{}$ are the SM-singlet right-handed neutrinos, while $N_{L,R}^{}$ are the other SM-singlet fermions and their mass matrix $\hat{m}_N^{}$ has been chosen real and diagonal without loss of generality and for convenience. Moreover, the model is protected from the other gauge-invariant terms including the Majorana masses of the SM-singlet fermions as well as the Yukawa couplings of the right-handed neutrinos to the SM.

After the electroweak symmetry breaking, the SM left-handed neutrinos $\nu_L^{}$ can obtain a Dirac mass term with the SM-singlet fermions $N_R^{}$, i.e.
\begin{eqnarray}
m_L^{}= y\langle\phi\rangle\,.
\end{eqnarray}
We then can conveniently express the masses of the neutral fermions $\nu_{L,R}^{}$ and $N_{L,R}^{}$ by a matrix, 
\begin{eqnarray}
\label{matrices}
\mathcal{L}&\supset& -\left[\bar{\nu}_L^{}~~\bar{N}_L^{}\right]\left[\begin{array}{cc}0& m_L^{}\\
[2mm]
m_R^{}& \hat{m}_N^{}\end{array}\right]\left[\begin{array}{c}\nu_R^{}\\
[2mm]
N_R^{}\end{array}\right] +\textrm{H.c.}\,. 
\end{eqnarray}
If the off-diagonal blocks $m_{L,R}^{}$ are much smaller than the diagonal block $\hat{m}_N^{}$, the left- and right-handed neutrinos $\nu_{L,R}^{}$ can acquire a tiny Dirac mass matrix, 
\begin{eqnarray}
\label{mass}
\mathcal{L} &\supset& - m_\nu^{} \bar{\nu}_L^{}\nu_R^{} +\textrm{H.c.} ~~\textrm{with}\nonumber\\
&&m_\nu^{} = -m_L^{} \frac{1}{\hat{m}_N^{}}m_R^{}\ll m_{L,R}^{}\ll m_N^{}\,,
\end{eqnarray}
while the SM-singlet fermions $N_{L,R}^{}$ can approximate to the heavy mass eigenstates,  
\begin{eqnarray}
N=N_L^{}+N_R^{}\,.
\end{eqnarray}
In the following we shall refer to the Dirac fermions $N$ as the mediator Dirac fermions since their mediation is crucial to the Dirac neutrino mass generation. 

In Eqs. (\ref{matrices}) and (\ref{mass}), the mass matrices $m_{L,R}^{}$ can be assigned to three possible hierarchies,
\begin{eqnarray}
(\textrm{i}) ~m_L^{} \sim m_R^{}\,, ~~(\textrm{ii}) ~m_L^{} \ll m_R^{}\,, ~~ (\textrm{iii})~ m_L^{} \gg m_R^{}\,. 
\end{eqnarray}
As we will clarify later, the case with $m_L^{} \gg m_R^{}$ can accommodate rich observable phenomena. This parameter choice is quite undemocratic. The neutrino mass generation in this case thus may be named as an undemocratic Dirac seesaw.

\section{Realistic models}

We now introduce a natural way to realize the undemocratic Dirac seesaw. Specifically, a new symmetry $S_\nu^{}$ is imposed to forbid the Yukawa couplings of the right-handed neutrinos to the SM lepton and Higgs doublets. Accordingly, the $\nu_R^{}-N_L^{}$ mass term $m_R^{}$ in Eq. (\ref{lag}) is expected to be induced through certain Yukawa interactions. The Lagrangian relevant to our demonstration is  
\begin{eqnarray}
\label{lag2}
\mathcal{L}&\supset& -y_{}^{} \bar{l}_{L}^{} \phi N_R^{} - \hat{m}_N^{} \bar{N}_L^{} N_R^{} - f_{a}^{} \sigma_a^{} \bar{N}_L^{} \nu_R^{} - \mu_{a}^{} \sigma^\dagger_{a} \xi^2_{} \nonumber\\
&&+\textrm{H.c.} - M_{a}^2 \sigma^\dagger_{a}\sigma_a^{}\,,~~(a=1,...,n\geq 2)\,. 
\end{eqnarray}
Here $\xi$ is a Higgs singlet responsible for spontaneously breaking the $S_\nu^{}$ symmetry, while $\sigma$ denotes some heavy Higgs singlets.

The new symmetry $S_\nu^{}$ can be a discrete, global or gauge symmetry. The discrete and global symmetries can be easily found. In the following, we shall briefly introduce two types of gauge symmetries,
\begin{itemize}
\item $U(1)_{B-L}^{}$ gauge symmetry for baryon and lepton number \cite{langacker2009},
\item $U(1)_{X}^{}$ gauge symmetry for dark photon \cite{okun1982,holdom1986,fh1991}.
\end{itemize}

For the $U(1)_{B-L}^{}$ symmetry, the simplest choice seems to consider three chiral fermions $(\nu_{R1}^{},\nu_{R2}^{},\zeta_{R}^{})$ with the $B-L$ numbers $(-4,-4,+5)$ \cite{mp2007,ms2015}. Then the two right-handed neutrinos $\nu_{R1,2}^{}$ can participate in the neutrino mass generation while the third chiral fermion $\zeta_{R}^{}$ can keep massless. More chiral fermions are also possible \cite{gu2019-3,gu2019-7,gu2019-8,pry2016,bbn2019}. For example, we can introduce four chiral fermions $(\nu_{R1}^{},\nu_{R2}^{},\chi_{R1}^{},\chi_{R2}^{})$ with the $B-L$ numbers $(-5/3,-5/3,-7/3,+8/3)$. When the two right-handed neutrinos $\nu_{R1,2}^{}$ contribute to the Dirac neutrinos, the other two chiral fermions $\chi_{R1,2}^{}$ can form a stable Dirac fermion to be a dark matter through their Yukawa coupling to the $U(1)_{B-L}^{}$ Higgs singlet $\xi$ which has a $B-L$ number $+1/3$, i.e. 
\begin{eqnarray}
\label{lag3}
\mathcal{L}\supset -y_\chi^{}\left(\xi\bar{\chi}_{R1}^{}\chi_{R2}^{c}+\textrm{H.c.}\right)\,.
\end{eqnarray}

In the $U(1)_X^{}$ case, the right-handed neutrinos $\nu_R^{}$ carry an opposite $X$ number to the heavy Higgs singlets $\sigma$. For example, we can consider five chiral fermions $(\nu_{R1}^{},\nu_{R2}^{}, \chi_{R1}^{}, \chi_{R2}^{},\zeta_{R}^{})$ with the $X$ numbers $(+1,+1,(1-\sqrt{145})/4,(1+\sqrt{145})/4,-5/2)$ when the $U(1)_X^{}$ Higgs singlet $\xi$ carries a $X$ number $-1/2$. We can also consider the possibility of three right-handed neutrinos. The $U(1)_X^{}$ Higgs singlet $\xi$ and the three right-handed neutrinos $\nu_{R1,2,3}^{}$ are still assigned to the $X$ number $-1/2$ and $+1$, respectively. The other three chiral fermions $(\chi_{R1}^{}, \chi_{R2}^{},\zeta_{R}^{})$ then can carry the $X$ numbers $((1-5\sqrt{17})/4,(1+5\sqrt{17})/4,-7/2)$. Besides the massless $\zeta_R^{}$ and the light $\nu_{R1,2}^{}$ or $\nu_{R1,2,3}^{}$, the other two chiral fermions $\chi_{R1,2}^{}$ can become a stable Dirac fermionic dark matter due to their Yukawa coupling with the $U(1)_X^{}$ Higgs singlet $\xi$, i.e. 
\begin{eqnarray}
\label{lag4}
\mathcal{L}\supset -y_\chi^{}\left(\xi\bar{\chi}_{R1}^c\chi_{R2}^{}+\textrm{H.c.}\right)\,.
\end{eqnarray}
Note in this $U(1)_{X}^{}$ case, we need impose a global symmetry of lepton number to forbid the Majorana masses of the SM-singlet fermions $N_{L,R}^{}$. However, this global symmetry will become unnecessary as well as the irrational X charges when we consider a left-right symmetric theory or a GUT. We will show shortly.

Obviously, the $U(1)_{B-L,X}^{}$ scenario can be directly embedded into an $SU(5)\times U(1)_{B-L,X}^{}$ framework since the $N_{L,R}^{}$ fermions are the $SU(5)$ singlets and hence their mass term $\hat{m}_N^{}$ is still gauge-invariant. However, if the $U(1)_X^{}$ scenario is expected to allow an $SU(3)_c^{}\times SU(2)_L^{} \times SU(2)_R^{} \times U(1)_{B-L}^{}$ left-right symmetric theory or an $SO(10)$ GUT, its assignment for the $U(1)_X^{}$ charges should be modified since the $N_R^{}$ fermions now belong to the $SU(2)_R^{}$ doublets or the $SO(10)$ 16-dimensional representations. At the left-right level, the first two terms in Lagrangian (\ref{lag2}) should come from  
\begin{eqnarray}
\label{lag5}
\mathcal{L}&\supset& -y_l^{}\bar{l}_L^{} \Phi l_R^{} - \tilde{y}_l^{} \bar{l}_L^{} \tilde{\Phi} l_R^{} - y_N^{}\left( \bar{l}_R^{} \eta_R^{} N_L^{} + \bar{l}_L^c \eta_L^\ast N_L^{}\right) \nonumber\\
&&+\textrm{H.c.}\,,
\end{eqnarray} 
where $l_{L,R}^{}$ are the lepton doublets, while $\eta_{L,R}^{}$ and $\Phi$ respectively are the Higgs doublets and bidoublet.

While the three generations of fermion multiplets do not carry any $X$ numbers, the three generations of the $N_L^{}$ fermions, the Higgs doublets $\eta_R^{}$ and $\eta_L^{}$ can be assigned to their $X$ numbers $-1$, $+1$ and $-1$, respectively. For the gauge anomaly cancellation, the simplest choice is to consider three chiral fermions $(\nu_{R1}^{},\nu_{R2}^{}, \zeta_{R}^{})$  with the X numbers $(-4,-4,+5)$. In this case, the $\zeta_R^{}$ fermion without mass can not serve as a dark matter particle. A more appealing scheme is to introduce more chiral fermions such as $(\nu_{R1}^{},\nu_{R2}^{}, \chi_{R1}^{},\chi_{R2}^{})$ with the $X$ numbers $(-5/3,-5/3,-7/3,+8/3)$. Accordingly, the Higgs singlets $\sigma_a^{}$ and $\xi$ should carry the $X$ numbers $+2/3$ and $+1/3$, respectively. Then the mass term $\hat{m}_N^{}$ in Eq. (\ref{lag2}) can be induced when the Higgs doublet $\eta_R^{}$ develops its vacuum expectation value (VEV) $\langle\eta_R^{}\rangle$ for the spontaneous left-right symmetry breaking. Meanwhile, the $\chi_{R1,2}^{}$ fermions can have a Yukawa coupling (\ref{lag3}) to form a stable Dirac fermion. Because the renormalizable terms like $\mu_{\eta\Phi}^{}\eta_L^\dagger \Phi \eta_R^{} $ and $\kappa_{\xi\eta\Phi}^{}\xi \eta_L^\dagger \Phi \eta_R^{}$ are forbidden by the $U(1)_X^{}$ symmetry, the Higgs doublet $\eta_L^{}$ can not obtain a nonzero VEV, however, it can contribute to a loop-level $U(1)$ kinetic mixing even if a tree-level $U(1)$ kinetic mixing is absent in the $SO(10)\times U(1)_X^{}$ framework. This means the $U(1)_X^{}$ gauge symmetry can provide a dark photon.

When we consider the $SO(10)$ GUT, the left-right symmetry should be broken at a very high scale. The mass term $\hat{m}_N^{}$ in Eq. (\ref{lag2}) thus can not be near the electroweak scale unless the related Yukawa couplings are extremely small. This problem can be solved as long as the Higgs doublet $\eta_R^{}$ is not responsible for the spontaneous left-right symmetry breaking. For this purpose, we can introduce the Higgs doublets $\omega_{R}$ and $\omega_L^{}$ with the $X$ numbers $+2/3$ and $-2/3$. The VEV $\langle\eta_R^{}\rangle$ thus can be induced through the cubic terms $\mu_{\xi\eta\omega}^{}\xi(\eta_R^\dagger \omega_R^{} + \eta_L^T \omega_L^\ast)+\textrm{H.c.}$. Meanwhile, the Higgs doublet $\omega_L^{}$ and hence the Higgs doublet $\eta_L^{}$ can not obtain their VEVs in the absence of the renormalizable terms like $\mu_{\omega\Phi}^{}\omega_L^\dagger \Phi \omega_R^{} $ and $\kappa_{\xi\omega\Phi}^{}\xi \omega_L^\dagger \Phi \omega_R^{}$.

\section{Neutrino mass and baryon asymmetry}

\begin{figure}
\centering
\includegraphics[scale=0.8]{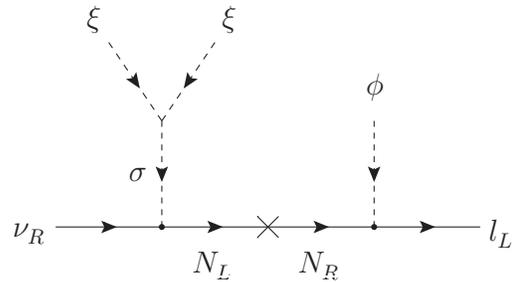}
\caption{The Dirac neutrino mass generation.}
\label{numass}
\end{figure}

After the Higgs singlet $\xi$ develops its VEV for spontaneously breaking the $S_\nu^{}$ symmetry, the heavy Higgs singlets $\sigma$ can pick up their suppressed VEVs, like the conventional type-II seesaw \cite{mw1980,sv1980,cl1980,lsw1981,ms1981}, i.e.
\begin{eqnarray}
\langle \sigma_a^{} \rangle \simeq -\frac{\mu_a^{}  \langle \xi\rangle^2_{}}{M_a^2} \ll \langle \xi \rangle ~~\textrm{for}~~M_a^{} \gtrsim \mu_a^{}\,, ~~M_a^{} \gg \langle\xi\rangle \,. ~~~~
\end{eqnarray}
The $\nu_R^{}-N_L^{}$ mass term $m_R^{}$ in Eq. (\ref{lag}) then can be given by 
\begin{eqnarray}
m_R^{} = \sum_a^{}f_a^{} \langle\sigma_a^{}\rangle\,.
\end{eqnarray}
The VEV $\langle\sigma\rangle$ can be expected to be much smaller than the electroweak one $\langle\phi\rangle$. In consequence, the undemocratic selection $m_L^{} \gg m_R^{}$ in the neutrino mass formula (\ref{mass}) can be elegantly explained. The complete neutrino mass generation can be understood in Fig. \ref{numass}.

\begin{figure*}
\centering
\includegraphics[scale=0.8]{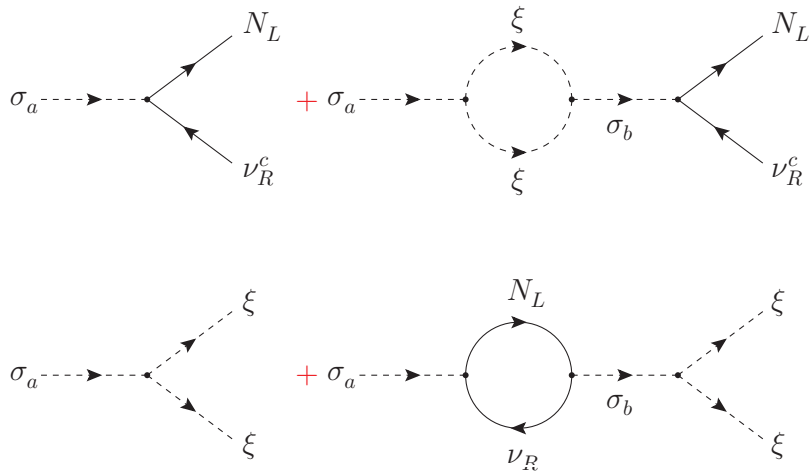}
\caption{The lepton-number-conserving decays of the heavy Higgs singlets. The lepton asymmetry stored in the fermions $N_L^{}$ will be partially converted to a baryon asymmetry by the sphaleron processes. The opposite lepton asymmetry stored in the right-handed neutrinos $\nu_R^{}$ (and probably also in the Higgs singlet $\xi$, depending on the definition of the global lepton number) will not participate in the sphaleron processes. }
\label{sdecay}
\end{figure*}

As shown in Fig. \ref{sdecay}, the heavy Higgs singlets $\sigma$ have two decay modes,
\begin{eqnarray}
\sigma \rightarrow  N_L^{}+\nu_R^c\,,~~ \sigma \rightarrow \xi +\xi \,.
\end{eqnarray}
If the CP is not conserved, we can expect a CP asymmetry in the above decays,
\begin{eqnarray}
\varepsilon_{\sigma_a}^{}&=& \frac{\Gamma(\sigma_a^{} \rightarrow N_L^{}+\nu_R^{c}  )-\Gamma( \sigma_a^{\ast} \rightarrow  N_L^{c}+\nu_R^{}  )}{\Gamma_{\sigma_a}^{}}\nonumber\\
&=&-\frac{\Gamma( \sigma^{}_{a} \rightarrow \xi+\xi)-\Gamma( \sigma_a^{\ast} \rightarrow \xi^\ast_{}+\xi^\ast_{})}{\Gamma_{\sigma_a}^{}}\nonumber\\
&=&-\frac{1}{2\pi}\sum_{b\neq a}^{}\frac{\textrm{Im}\left[\textrm{Tr}\left(f_a^\dagger f_b^{} \right) \mu_a^{}\mu_b^\ast\right]}{\textrm{Tr}\left(f_{a}^{\dagger}f_a^{}\right)+\frac{2|\mu_a^{}|^2_{}}{M_{a}^{2}}}\frac{1}{M_b^2 -M_a^2}\,,
\end{eqnarray}
where $\Gamma_{\sigma_a}^{}$ is the total decay width,
\begin{eqnarray}
\Gamma_{\sigma_a}^{}&=&\Gamma(\sigma_a^{} \rightarrow N_L^{}+\nu_R^{c})  + \Gamma( \sigma_a^{} \rightarrow \xi +\xi )\nonumber\\
&=&\Gamma( \sigma_a^{\ast} \rightarrow  N_L^{c}+\nu_R^{}  )+\Gamma( \sigma^\ast_{a} \rightarrow \xi^\ast_{}+\xi^\ast_{})\nonumber\\
&=&\frac{1}{16\pi}\left[\textrm{Tr}\left(f_{a}^{\dagger}f_a^{}\right)+\frac{2|\mu_a^{}|^2_{}}{M_{a}^{2}}\right]M_{a}^{}\,.
\end{eqnarray}
It should be noted that a nonzero CP asymmetry $\varepsilon_{\sigma_a}^{}$ needs at least two heavy Higgs singlets $\sigma_{a,b\neq a}^{}$.

After the heavy Higgs singlets $\sigma_a^{}$ go out of equilibrium, their decays can generate a lepton symmetry $L_{N}^{}$ stored in the fermions $N_L^{}$, i.e.
\begin{eqnarray}
\label{xasymmetry}
L_N^{}&=& \sum_{a}^{}\frac{\varepsilon_{\sigma_a}^{}}{g_\ast^{}} \kappa_a^{}\,.
\end{eqnarray}
Here the symbol $g_\ast^{}$ is the relativistic degrees of freedom, while the factor $\kappa_a^{}<1$ represents the effect of washout processes \cite{kt1990}. In order to determine the washout factor $\kappa_a^{}$ we can solve the full Boltzmann equations. The $L_N^{}$ asymmetry can be partially converted to a baryon asymmetry $B$ by the sphaleron processes, 
\begin{eqnarray}
\label{sph}
B= c L_{N}^{}\,,
\end{eqnarray}
where the factor $c$ depends on the masses and probably also the number of the Dirac fermions $N$. For example, we can obtain $c=-28/79$ \cite{ht1990} if the lepton asymmetry $L_N^{}$ is produced before the sphalrons \cite{bcst1999}, meanwhile, the Dirac fermions $N$ are heavy enough and hence have completely decayed before the electroweak symmetry breaking.

\section{Phenomenology}

In the undemocratic Dirac seesaw, the mediator Dirac fermions $N$ can be relatively light but significantly mix with the left-handed neutrinos $\nu_L^{}$. The induced non-unitarity effect in the lepton mixing matrix can have an important impact on the neutrino oscillations \cite{abfgl2006,efmtv2015,mtv2016}. The sizable $N-\nu_L^{}$ mixing can also lead to some charged lepton flavor violating processes such as $\mu\rightarrow e \gamma$, $\mu\rightarrow 3e$, and $\mu \rightarrow e$ conversion in nuclei \cite{bsvmv1987,dkv2005,rhms2016}.  

The mediator Dirac fermions $N$ without too heavy masses can significantly contribute to the invisible Higgs decays through their Yukawa couplings with the left-handed neutrinos $\nu_L^{}$. The collider implications of invisibly decaying Higgs boson have been extensively discussed by theorists \cite{cerv1997} and also studied by the LHC collaborations \cite{sirunyan2018,aaboud2019}. The future collider experiments such as the CEPC \cite{cepc2018} will be also interested in the invisible Higgs decays.

The mediator Dirac fermions $N$ with an accessible mass can be produced at colliders due to their sizable mixing with the left-handed neutrinos $\nu_L^{}$ \cite{dsgv1990,mr2016}. The $U(1)_{B-L}^{}$ gauge boson or the $U(1)_X^{}$ dark photon can also enhance the production cross section of the mediator Dirac fermions $N$ at colliders \cite{ddkv2012,hhhl2018,jllpsy2019}.

Remarkably, ones may have noticed that the above phenomena from the mediator Dirac fermions in our undemocratic Dirac seesaw actually are very similar to those from the mediator quasi-Dirac fermions in the conventional inverse seesaw \cite{mv1986,rsv2019}. In other words, it seems difficult to distinguish our undemocratic Dirac seesaw and the conventional inverse seesaw by ongoing and planned experiments. 

The $U(1)_{B-L}^{}$ gauge boson or the $U(1)_X^{}$ dark photon can decay into the right-handed neutrinos $\nu_R^{}$, the massless fermion $\zeta_R^{}$, and probably also the Dirac fermionic dark matter $\chi$ and the mediator Dirac fermions $N$. The running and planned collider experiments can be expected to verify the $U(1)_{B-L}^{}$ gauge boson \cite{afpr2017,klq2016} or the $U(1)_X^{}$ dark photon \cite{hhhl2018,jllpsy2019}.

The $U(1)_{B-L}^{}$ gauge boson or the $U(1)_X^{}$ dark photon can dominate the decoupling of the massless fermion $\zeta_R^{}$ and the right-handed neutrinos $\nu_{R}^{}$ \cite{gu2019-3,gu2019-4,gu2019-7,gu2019-8}, besides the annihilation and scattering of the Dirac fermionic dark matter $\chi$ \cite{gu2019-3,gu2019-4,gu2019-7,gu2019-8,gh2017}. The BBN constraint on the effective neutrino number requires the right-handed neutrinos $\nu_R^{}$ and the massless state $\zeta_R^{}$ should decouple above the QCD scale \cite{kt1990}.

\section{Conclusion}

In this paper we have proposed an undemocratic Dirac seesaw mechanism to simultaneously generate the small neutrino masses and the cosmic baryon asymmetry. In our scenario, the mediator Dirac fermions can have a sizable mixing with the left-handed neutrinos and a suppressed mixing with the right-handed neutrinos. Therefore, the mediator Dirac fermions can be relatively light and hence result in rich observable phenomena. The right-handed neutrinos without any Yukawa couplings to the SM lepton and Higgs doublets can be naturally introduced by the $U(1)_{B-L}^{}$ gauge symmetry for baryon and lepton number or the $U(1)_X^{}$ gauge symmetry for dark photon. The $U(1)_{B-L}^{}$ or $U(1)_X^{}$ gauge symmetry can also predict the existence and guarantee the stability of the Dirac fermionic dark matter. The $U(1)_{B-L}^{}$ gauge boson or the $U(1)_X^{}$ dark photon can also lead to some interesting phenomena including the searches for the dark matter and the mediator Dirac fermions. Our scenario can be naturally embedded into the left-right symmetric theory or the GUT.

\textbf{Acknowledgement}: This work was supported in part by the National Natural Science Foundation of China under Grant No. 12175038 and in part by the Fundamental Research Funds for the Central Universities.

\end{document}